 \documentclass[apj]{emulateapj}

\usepackage{graphicx,hyperref}
\usepackage{txfonts}
\usepackage{epstopdf}
\usepackage{longtable}
\usepackage{natbib}
\usepackage{color}

\newcommand{\kms}{{km\,s}$^{-1}$}

\newcommand{\teff}{$T_\mathrm{eff}$\,}
\newcommand{\logg}{$\log g$\,}

\newcommand{\vsini}{$v \sin i$}

\newcommand{\vt}{$v_{\rm{t}}$}

\shorttitle{B-type Galactic Runaways}
\shortauthors{C. M. McEvoy et al.}

\begin{document}
\title{The origin of B-type runaway stars: \\ Non-LTE abundances as a diagnostic}


\author{Catherine M. McEvoy \altaffilmark{1,2,3}, Philip L. Dufton \altaffilmark{1}, Jonathan V. Smoker \altaffilmark{2,1}, David L. Lambert \altaffilmark{4}, Francis P. Keenan \altaffilmark{1}, Fabian R. N. Schneider \altaffilmark{5}, Willem-Jan de Wit \altaffilmark{2}}

\altaffiltext{1}{Astrophysics Research Centre, School of Mathematics and Physics, Queen's University Belfast, Belfast BT7 1NN, UK}

\altaffiltext{2}{European Southern Observatory, Alonso de Cordova 3107, Casilla 19001, Vitacura, Santiago 19, Chile}

\altaffiltext{3}{Graduate School, King's College London, London SE1 9NH, UK}

\altaffiltext{4}{The University of Texas at Austin,
Department of Astronomy,
RLM 16.316,
Austin, TX 78712}

\altaffiltext{5}{Department of Physics,
University of Oxford,
Denys Wilkinson Building, Keble Road,
Oxford
OX1 3RH,
UK}

 \voffset=-17.5mm

\begin{abstract}

There are two accepted  mechanisms to explain the origin of runaway OB-type stars: the Binary Supernova Scenario (BSS), and the Cluster Ejection Scenario (CES). In the former, a supernova explosion within a close binary ejects the secondary star, while in the latter close multi-body interactions in a dense cluster cause one or more of the stars to be ejected from the region at high velocity.  Both mechanisms have the potential to affect the surface composition of the runaway star.
TLUSTY non-LTE model atmosphere calculations have been used to determine atmospheric parameters and carbon, nitrogen, magnesium and silicon abundances for a sample of B-type runaways. These same analytical tools were used by \cite{hun09a} for their analysis of 50 B-type open cluster Galactic stars (i.e. non-runaways).  Effective temperatures were deduced using the silicon-ionization balance technique, surface gravities from Balmer line profiles and microturbulent velocities derived using the Si spectrum.  The runaways show no obvious abundance
anomalies  when compared with stars in the open clusters. The runaways do show a spread in composition which almost certainly reflects the Galactic abundance gradient and a range in the birthplaces of the runaways in the Galactic disk.  Since the observed Galactic abundance gradients of C, N, Mg and Si are of a similar magnitude, the
abundance ratios (e.g., N/Mg) are, as obtained, essentially uniform across the sample. 
\end{abstract}


%
\section{Introduction}                                         \label{s_intro}
 
The presence of a significant number of early-type OB main sequence
stars in the Galactic halo, as  demonstrated by \citet{green74}, 
has occasioned an extensive literature on the origins of 
massive young stars far from the Galactic disk, the richest site of 
star-forming regions. Very broadly, two explanations for halo OB main
sequence stars have survived scrutiny; both  explanations  consider
that the stars formed in the Galactic disk and were
ejected from their parental open cluster or association with sufficent
velocity to reach the Galactic halo. 
Then, the stars deserve their common classification as `runaway' stars.
A less likely explanation not considered further here is that
the halo OB main sequence stars were formed {\it in situ} \citep{dyson1983}.

Two proposed scenarios are considered capable of producing
runaway stars: the binary supernova scenario (here, BSS)  and the
cluster ejection scenario (here, CES): \\ 
\noindent -- In the BSS proposed first
by \citet{zwicky1957} and developed by \citet{blaauw1961}, the runaway star 
was the secondary in a binary with a more massive star which experienced
its terminal supernova explosion. 
The reduced gravitational attraction of the primary's 
stellar remnant, either a neutron star or a black hole,  freed the
secondary to escape with a velocity similar to its orbital velocity. The escaping secondary - the runaway star -
may be accompanied by the stellar remnant. \\
\noindent -- In the CES proposed by \citet{poveda1967}, close encounters
in a young open cluster may lead to ejection of a star. Multi-body
interactions are favored to eject a single or a binary star. Simulations
suggest that the most effective way to produce high-velocity runaway
stars is through the interaction of two hard binary systems \citep{hoffer1983}.
\citet{leonard1989} and \citet{leofahl1991} show that binary-binary
interactions may result in runaway single, binary or even merged binary stars.

As might be expected, both BSS and CES contribute to the runaway population.
Hoogerwerf et al. (2001 - see also \citealt{dez2001}) use astrometric
data to predict past tracks of stars in the Galaxy to show 
that specific examples of runaways may be attributed to
the BSS (see $\zeta$ Oph and pulsar PSR J1392+1059) or to the CES 
(AE Aur and $\mu$ Col ejected from Orion -- see \citet{blamor1954} and
\citet{gies1986}.  Across their sample of runaways,
Hoogerwerf et al. estimate that two-thirds arise from the BSS and one-third
from the CES.  Other authors consider the CES the greater contributor of runaway stars.

In this paper, we provide a non-LTE analysis of C, N, Mg and Si abundances for a sample of runaway
B stars and search for abundance differences among the sample and between the
sample and B stars in young open clusters in the Galactic disk. The goal is
to determine, if the abundance information provides convincing evidence
or even intriguing clues to the origin of a runaway. The suggestion to exploit chemical
composition to judge  competing origins of runaway stars is
traceable to \citet{blaauw1993} who suggested a study of the He/H abundance
versus projected rotational velocity $v\sin i$ with He enrichment and high
$v\sin i$ resulting from the BSS. 
 Helium enrichment should generally be accompanied
by N enrichment and a parallel C deficiency as a result of H-burning by the
CNO-cycles. A Mg and Si enrichment of the runaway star might result for
stars provided by the BSS but not the CES.  
Many previous studies have reported LTE
abundance analyses generally giving abundances relative to Galactic disk B stars of the same
atmospheric parameters - see, for example, the runaway sample analyzed by \citet{martin2004}. This is the
first non-LTE analysis of a sample of runaway B stars.

In Section 2, we discuss selection criteria used to isolate our sample of
runaway stars whose spectra were obtained (see Section 3) and analyzed
in Section 4
following the method previously applied by Hunter et al. (2009 - see also
\citealt{trun2007}) to B stars in the three
Galactic open clusters. Abundances in the runaway and cluster B stars
are discussed in Section 5.  Rotational and radial velocities are discussed in Section 6.
Brief  concluding remarks are  offered in Section 7.

\section{Selection of Runaway B stars} \label{s_sel} 

Targets were selected from catalogues of previously identified runaway candidates. Different criteria are outlined in each source identified in the 
reference column of Table~\ref{stars_info} and cited in the footnotes. All objects in our sample either lie far from the Galactic plane with a height above or below the plane (z distance) of $> 0.3$ kpc or have high Galactic latitude ($|b| > 30 ^{\circ}$) or a peculiar space velocity of $\gtrsim 30$ \kms. A star that meets any one of these criteria is considered  a runaway.  
 Essentially each star in our sample has been shown by
one or more of the references to have been ejected from the disk, i.e., the travel time from disk to its halo
location is less than the lifetime of the star.
In the BSS it is expected that the ejection velocity gained from the loss of mass from the binary system as a result
of the supernova will  not exceed 300-400\kms, while for CES, similar velocities are possible \citep[see][]{leo93, por00, gva09}. A majority of the runaway stars are expected to be bound to the Galaxy. Stars exceeding the escape
velocity are generally called `hypervelocity' stars and, the principal ejection engine for such stars is the Galaxy's
central supermassive blackhole  \citep[see review by][]{brown15}.  Two hypervelocity stars originating in the outer Galaxy are
known: HD 271791 \citep{heber08,prz08} and HIP 60350 \citep{irr10}. Such stars may be
products of the BSS operating in a binary having  particular initial masses.

One of our two primary sources of runaway B stars is a list 
 by \citet{silva2011}
of 174 high Galactic latitude B stars drawn from the literature.
As noted by Silva \& Napiwotzki and emphasized by essentially all previous discussion of runaway B stars, spectral classification of B-type
does not ensure that the star is a B main sequence
(massive) star because low mass stars evolving either off the blue horizontal
branch or from the asymptotic branch stars (i.e., post-AGB stars) can encroach on the effective
temperature - surface gravity plane $(T_{\rm eff},\log g)$ 
plane occupied by main sequence B stars -
see also \citet{tobin1987}. Using then available
information on  $(T_{\rm eff},\log g)$ and
 data on atmospheric abundances, Silva \& Napiwotzki identified
which of the 174 stars belong or possibly belong to the main sequence
(MS or MS?). With one exception, the  stars we observed from 
Silva \& Napiwotzki's Tables 2 and 3 were classified as MS or MS? The
one exception is HIP 60615 (BD +36 2268) which was classified Non-MS.  As massive
stars, the combination of effective temperature and surface gravity indicate that the
sample spans the mass range of about 5--25 solar masses.  
Estimated ejection velocities range up to about 400\kms \citep{silva2011} and do not lead to escape from the
Galaxy. Actual space velocities of observed runaways are smaller than the ejection velocities because stars are generally observed
near  the apex of their orbit where they spend most of their time.

Our second primary source of runaway B stars is the catalogue 
of young runaway stars within 3 kpc of the Sun
compiled by \citet{tetz2011} using {\it Hipparcos}
astrometry. From this catalogue of more than 2500 stars younger than 
about 50 My and with peculiar velocities, 
we selected 13 B stars and of these six belong also to
our selection from Silva \& Napiwotzki. 

Our sample was completed by selecting another
10 targets from the literature on runaway stars.

References to all sources providing runaway stars  are noted in Table~\ref{stars_info} listing 38 stars. The table lists the stars by  their HIP number where available and/or an alternative designation, the V magnitude, the spectral type, the source of our spectra (see next Section), the radial velocity ($v_r$), the projected rotational velocity ($v\sin i$) and the reference to the star's selection as a runaway B star.


\begin{table*} 
\caption{Each star listed by HIP number, along with an alternative identifier. V magnitudes, spectral types, instrument used for observation, radial velocity, projected rotational velocity and the reference that identifies the star as a runaway. }  
\label{stars_info}
\centering
\begin{tabular}{cccccccc}   \hline
 \multicolumn{1}{c}{HIP} & \multicolumn{1}{c}{Other} & \multicolumn{1}{c}{Vmag} & \multicolumn{1}{c}{Spec.} & 
 \multicolumn{1}{c}{Obs.$^{a}$} & \multicolumn{1}{c}{$v_r$} & \multicolumn{1}{c}{$v\sin i$} & \multicolumn{1}{c}{Ref$^{b}$}   \\ 
 \multicolumn{1}{c}{ } &  \multicolumn{1}{c}{} & \multicolumn{1}{c}{} & \multicolumn{1}{c}{type} & \multicolumn{1}{c}{} & 
 \multicolumn{1}{c}{km-s$^{-1}$} &  \multicolumn{1}{c}{km-s$^{-1}$} &  \multicolumn{1}{c}{}  \\ \hline

  2702 &  HD 3175    &   9.33 & B4V     & F  &  -13$\pm$2 &   26$\pm$2 & S11  \\
  3812 &CD -56 152   &  10.18 & B2V     & U  &   14$\pm$8 &  194$\pm$7 & S11  \\
  7873 & HD 10747    &   8.15 & B2V     & F  &   -9$\pm$2 &   15$\pm$1 & T11  \\
 13489 & HD 18100    &   8.44 & B5II/III & F &   80$\pm$7 &  241$\pm$6 & M05  \\
 16758 & HD 22586    &   9.33 & B4V     & F  &  -13$\pm$2 &   88$\pm$3 & S11  \\
 45563 & HD 78584    &   8.20 & B3      & T  & -120$\pm$6 &  102$\pm$4 & T11  \\
 55051 & HD 97991    &   7.41 & B2/3V   & U  &   31$\pm$3 &  135$\pm$3 & S11  \\
 56322 & HD 100340   &  10.12 & B0      & T,U&  253$\pm$10&  181$\pm$10 & S11  \\
 60615 &BD +36 2268  &  10.31 & B3V     & T  &   31$\pm$4 &   54$\pm$4 & S11  \\
 61431 & HD 109399   &   7.67 & B0.5III & F  &  -43$\pm$3 &  203$\pm$6 & T11  \\
 64458 & HD 114569   &   8.10 & B7/8    & F  &  104$\pm$2 &   74$\pm$1 & M12  \\
 67060 & HD 119608   &   7.53 & B1Ib    & F  &   31$\pm$1 &   59$\pm$9 & M04  \\
 68297 & HD 121968   &  10.26 & B1V     & T,U&   17$\pm$9 &  199$\pm$27 & S11 \\
 70205 & LP 857-24   &  11.36 &$\ldots$ & F  &  243$\pm$4 &   64$\pm$4 & B12  \\
 70275 & HD 125924   &   9.66 & B2IV    & T  &  244$\pm$1 &   64$\pm$3 & S11  \\
 79649 & HD 146813   &   9.06 & B1.5    & T  &   21$\pm$2 &   87$\pm$3 & S11  \\
 81153 & HD 149363   &   7.81 & B0.5III & F,T,U & 145$\pm$3 & 88$\pm$10 & S11 \\
 85729 & HD 158243   &   8.15 & B1Ib    & F  &  -63$\pm$2 &   70$\pm$2 & M12  \\
 91049 & HD 171871   &   7.78 & B2IIp   & T  &  -64$\pm$1 &   44$\pm$1 & T11  \\
 92152 & HD 173502   &   9.70 & B1II    & F  &   49$\pm$1 &   53$\pm$3 & K82  \\

%

 94407 & HD 179407   &   9.44 & B0.5Ib  & F  & -120$\pm$4 &  134$\pm$9 & S97  \\
 96130 & HD 183899   &   9.93 & B2III   & F  &  -46$\pm$2 &   55$\pm$3 & S11, T11  \\
 98136 & HD 188618   &   9.38 & B2II    & F  &   46$\pm$4 &  167$\pm$3 & S11  \\
101328 & HD 195455   &   9.20 & B0.5III & F,U &  19$\pm$7 &  213$\pm$7 & S11  \\
105912 & HD 204076   &   8.73 & B1V     & F,U &   0$\pm$3 &  102$\pm$1 & S11, T11 \\
107027 & HD 206144   &   9.34 & B2II    & F,U & 122$\pm$5 &  184$\pm$12 & S11  \\
109051 & HD 209684   &   9.94 & B2/3III & U   &  82$\pm$2 &  108$\pm$3 & S11, T11 \\
111563 & HD 214080   &   6.93 & B1/2Ib  & F   &  16$\pm$2 &  108$\pm$3 & S11  \\
112022 & HD 214930   &   7.40 & B2IV    & T   & -60$\pm$4 &   12$\pm$1 & M05  \\
112482 & HD 215733   &   7.34 & B1II    & T   &  -6$\pm$6 &   72$\pm$1 & T11  \\
113735 & HD 217505   &   9.13 & B2III/IV& F   & -17$\pm$1 &   26$\pm$2 & S11  \\
114690 & HD 219188   &   7.06 & B0.5III & F,T &  73$\pm$19 & 239$\pm$15 & S11  \\
115347 & HD 220172   &   7.64 & B3Vn    & F   &  26$\pm$2  &  39$\pm$1 & S11  \\
115729 & HD 220787   &   8.29 & B3III   & F   &  26$\pm$2  &  26$\pm$2 & S11, T11 \\
$\ldots$&EC 05582-5816 & 9.46 & B3V     & F   &  85$\pm$13 &  221$\pm$5 & S11  \\
$\ldots$&EC 13139-1851 & 10.50 & B4     & U  &   15$\pm$4  &  43$\pm$1 & S11  \\
$\ldots$&EC 20140-6935 &  8.83 & B2V    & F  &  -24$\pm$2  &  45$\pm$2 & S11  \\
$\ldots$& PB 5418      & 11.35 & B2     & U  &  147$\pm$3  &  50$\pm$1 & S11  \\
$\ldots$& PHL 159      & 10.90 & B      & U  &   87$\pm$2  &  30$\pm$1 & S11 \\

\hline
\end{tabular}

\flushleft {
Note: $^{a}$ Spectrograph used for the observation: F=Feros, T=Tull and U=UVES \\
$^{b}$ References: S11=\citet{silva2011},  T11=\citet{tetz2011}, M05=\citet{mdzin2005},
    M12=\citet{mcd2012}, M04=\citet{martin2004}, B12= \citet{debru2012}, K82=\citet{keen1982}, S97=\citet{smartt1997}. 
           }
 \end{table*}


\section{Observations} \label{s_obs}

High-resolution high signal-to-noise optical spectra were collected from three telescopes over a period of 16 months. Wavelength coverage and observation dates of each dataset are shown in Table \ref{obs_coverage}.
Each set of observations is described below, along with the reduction procedures through which they were prepared for analysis. 

\subsection{FEROS Observations}

Twenty six targets were observed in August 2014, using the ESO - FEROS instrument \citep{kau99}, a high resolution (R $\approx$ 48,000) prism cross-dispersed echelle spectrograph. FEROS has almost complete spectral coverage from 3500--9200\,\AA\, and provides high SNR ($\approx$ 300 - see Table \ref{obs_coverage}) in relatively short exposure times for bright stars. All data were reduced  using the ESO FEROS pipeline (version  1.57). 
 Multiple exposures were combined using either a median or weighted average, within \textsc{iraf}\footnote{IRAF is distributed by the National Optical Astronomy Observatory, which is operated by the Association of Universities for Research in Astronomy (auRA) under cooperative agreement with the National Science Foundation.}. Both these methods of merging the exposures resulted in very similar spectra, with the weighted average giving a slightly higher signal-to-noise ratio, and so these were adopted. Occasionally cosmic ray events were visible, but these were easily removed from the spectra manually, if they interfered with any analysis. 

Four targets (HD 1999, HD 165955, HD 204076 and HD 208213) were deemed unusable for  abundance analysis. HD 1999 showed double lines in its spectrum, making quantitative analysis unreliable. HD 165955 has a very high \vsini\, and so many lines became unobservable. HD 204076 proved to be double-lined spectroscopic binary when observed with UVES. HD 208213 did not have a sufficient SNR to identify Si {\sc iii} lines required for our analysis.  It was also observed with UVES and discarded for the same reason.

 \begin{table}

\caption{Observation dates and wavelength coverage of each instrument} 
\label{obs_coverage}
\begin{center}
\begin{tabular}{lll}
\hline\hline
\multicolumn{1}{c}{Instrument} & \multicolumn{1}{c}{$\lambda$ coverage (\AA)} & \multicolumn{1}{c}{Dates} \\
\hline
FEROS & 3500 -- 9200 & August 2013\\
Tull & 3400 -- 10900 & May/June 2013\\
UVES blue arm & 3730 -- 4990 & March-July 2014\\
UVES red arm & 5650 -- 9460 &March--July 2014\\
\hline \hline
\end{tabular}
\end{center}

\end{table}

\subsection{UVES Observations} 

Seventeen targets were observed at the VLT using the UVES instrument \citep{dek00}, a high resolution (R $\approx$ 80,000), high efficiency, cross-dispersed echelle spectrograph with a blue and a red arm. 
A standard setting (437 + 760) was used, yielding a wavelength coverage of 3730 -- 4990 \,\AA~ in the blue arm and 5650 -- 9460\,\AA~ in the red. 
All of the data were taken directly from the ESO archive, having been reduced using the ESO UVES pipeline \citep{bal00}. Multiple exposures were normalised and combined using either a median or weighted $\sigma$-clipping algorithm, within IDL. As the SNR of the individual exposures was high, the final spectra from both methods were effectively indistinguishable. Again, any cosmic ray events that interfered with subsequent analysis were removed manually.

As the UVES dataset was obtained to extend our study to higher \vsini\, and fainter targets, some of these proved particularly difficult to analyze, and five were discarded leaving twelve for which atmospheric parameters and  abundances have been estimated. 

\subsection{McDonald Observations}\label{McD_Obs}

Fourteen runaway candidates were observed at the W. J. McDonald Observatory with the Tull coud\'{e} echelle spectrograph at the 2.7m Harlan Smith telescope \citep{tul95}, with a high spectral resolution (R $\approx$ 60,000), during May and June 2013. Spectra were obtained covering a wavelength range 3,400--10,900\,\AA. The echelle data were split into orders, each of which was reduced separately using standard \textsc{iraf} procedures. 

For stars with significant rotational broadening and/or large surface gravities, the H$\delta$ line profiles extended over a significant fraction of the order. Therefore, normalization of these data proved more difficult than those from UVES and FEROS. To deal with this, blaze fits were made to the bracketing orders, where the continuum was  obvious, and these were averaged to provide a blaze profile for the order containing H$\delta$. This was then used to rectify the orders containing H$\delta$. All data were subsequently normalized and multiple exposures combined using either a median or weighted $\sigma$-clipping algorithm, within IDL, as above. 

Three of the McDonald targets were not analyzed. HD~69686 and HD~118246 did not have sufficient SNR to identify the rotationally broadened Si {\sc iii} lines, preventing estimates of the projected rotational velocities, effective temperatures and microturbulences. HD~203664 had complex and variable spectra \citep{aer06} and so was removed from the sample.  


\section{Method of analysis}\label{s_meth}

Non-LTE model atmosphere grids and model atoms from the {\sc tlusty} and  {\sc synspec} codes \citep{hub88, hub95, hub98, lan07} were used to  derive atmospheric parameters and chemical abundances. More detailed discussions of our analysis methods can be found in \citet{ hun07}, while those of the atmospheric grids are in \citet{rya03} and \citet{duf05}\footnote{See also http://star.pst.qub.ac.uk}. Hence only a brief summary is given here.

Model atmosphere grids have been generated with metallicities representative of the Galaxy ([$\log$ {Fe}/{H})\, + 12] = 7.5 dex, and other abundances scaled accordingly). 
These model atmospheres cover a range of effective temperatures from 12\,000 K to 35\,000 K in steps of 500 -- 1500 K, and surface gravities ranging from close to the Eddington limit to 4.5 dex in steps of 0.15 dex  \citep{hun08b}. 

The codes make non-LTE assumptions,  i.e. the atmospheres can be considered plane parallel with winds having no significant effect on the optical spectrum, and a normal helium to hydrogen ratio (0.1 by number of atoms) was assumed. \citet{duf05} and \citet{mce15} independently tested the validity of this approach. They analysed the spectra of B-type supergiants in the SMC and LMC, respectively, using the grids described here and also the {\sc fastwind} code \citep{san97,pul05} which incorporates wind effects. \citet{duf05} found excellent agreement in the atmospheric parameters estimated from the two methods. Effective temperature, logarithmic surface gravity, and microturbulent velocity estimates all agreed to well within their errors. Abundance values agreed to within 0.1 dex for elements such as C, O, Si and Mg, while discrepancies in N abundances were less than 0.2 dex, although a systematic difference of $~$0.1 dex did appear to exist between the two approaches. \citet{duf05} suggested that this was due to differences in the N model atoms and wind effects adopted. \citet{mce15} also found good agreement between results from the two codes. Of the eleven stars analysed using both methods, five targets had effectively identical results, five agreed well (with differences $\le$ 1000 K in \teff, $\le$ 0.1 dex in \logg, and $\le$ 0.2 dex in N abundance estimates), with the N abundance in only one target showing significant discrepancies between the methods. However, this star is an extreme object, close to the Eddington limit, where N abundances will be less secure. As the majority of our sample are lower luminosity dwarfs or giants, our approach should be adequate.

We adopt baseline chemical abundances from \cite{hun07,hun09a} derived from a sample of B-type stars in the Milky Way. Other B-type stellar studies, such as those by \cite{lyu05, lyu13,daf09,sim10,nie11}, have found  higher baseline abundances for Mg, Si and N, closer to Solar abundances, but to maintain consistency in our analysis, we use the Hunter baseline values which for N, Si and Mg, are 7.62, 7.42 and 7.25 dex, respectively \citep[see][]{hun07,hun09a}.

It is important to note that the stellar metallicity distribution in a disk galaxy, including the Milky Way, typically exhibits a negative gradient as a function of distance from the Galactic centre, both in the radial and vertical directions \citep{hua15}. Examples of studies where radial N abundance gradients in the Galactic disk have been studied include \cite{rol00}, who found a gradient of $-0.09~\pm$ 0.01~{dex~kpc$^{-1}$}, while \cite{daf04} found an average gradient for all elements in the Galactic disk of  {$-0.042~\pm$~0.007~dex~kpc$^{-1}$}, with an N value of $-0.046~\pm~0.011$~dex~kpc$^{-1}$, half of that estimated by \cite{rol00}. Both of these measurements rely on abundances in young OB-type stars. \cite{sha83} used radio and optical spectroscopy to sample Galactic H~{\sc ii} regions, spanning 3.5--13.7~kpc from the Galactic centre. They found a N abundance gradient of $-0.09~\pm$~0.015~dex~kpc$^{-1}$, similar to \cite{rol00}, along with evidence of steeper gradients in the inner regions of the Galactic disk. \cite{hua15} investigated this effect both radially and vertically using 7000 Red Clump stars between 7 and 14~kpc from the Galactic centre. They found that between 7 and 11.5 kpc the radial gradient flattens as the height from the Galactic plane increases, but that between 11.5 and 14~kpc the gradients do not vary with height and are at a constant value of --0.014~dex~kpc$^{-1}$. \cite{rol00} also considered a two-zone model, but found no evidence to indicate this was more appropriate. Here, an average N abundance for Galactic B-type stars has been adopted from \cite{hun09a}, although this value may be slightly higher if stars were formed closer to the Galactic centre, and lower if formed far from it. However, in no case has the gradient been found to be very large, and so the metallicity distribution should not have a significant effect on the results presented here.  

\subsection{Atmospheric parameters} \label{s_parameters}

The three characteristic parameters of a static stellar atmosphere (effective temperature, surface gravity and microturbulence) are inter-dependent and so an iterative process was used to estimate these values \citep[see][for more details]{fra10, mce15}. The parameters are described separately below. The final values for these parameters are given in Table \ref{stars_atmos}. 


\begin{figure*}
\begin{center}
\includegraphics[trim=1.4cm 0.2cm 1.5cm 1.3cm, clip=true,height=0.45\textheight,width=0.98\textwidth]{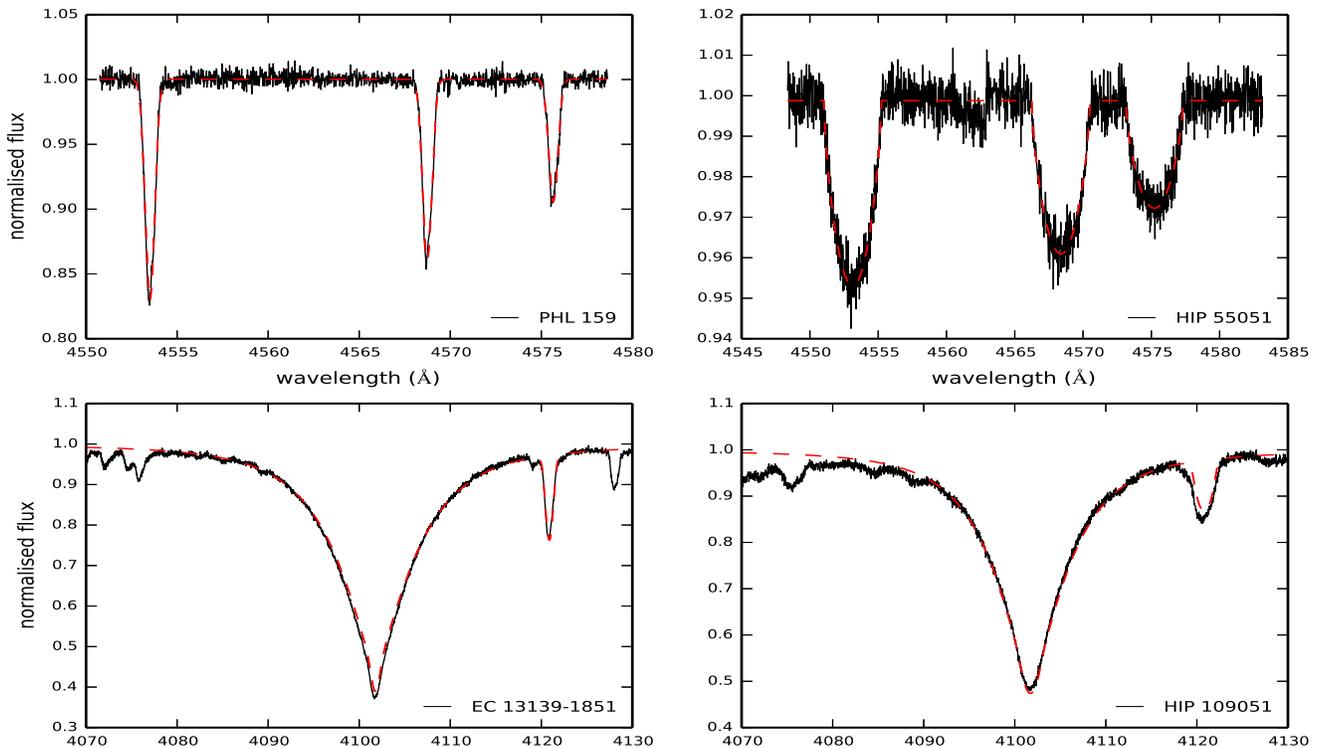}
\caption{Upper left: Si {\sc iii} spectrum for PHL 159 (\vsini\, = 30 \kms), with gaussian model fits (red dotted line) used to determine \vt\, and \teff.
Upper right: Si {\sc iii} spectrum for HIP 55051 (\vsini\, = 135 \kms), with rotationally broadened model fits (red dotted line) used to determine \vt\, and \teff.
Lower left: H $\delta$ spectrum for EC 13139-1851 (\vsini\,= 43 \kms), along with its model fit, used to determine \logg. 
Lower right: H $\delta$ spectrum for HIP 109051 (\vsini\,= 108 \kms), along with its model fit, used to determine \logg.}
 \label{fits}
\end{center}
\end{figure*}

\subsubsection{Effective Temperature}

Effective temperature (\teff) estimates were determined using the Si ionisation balance. Equivalent widths of the {Si} {\sc iii} multiplet (4552, 4567, 4574\,\AA) were measured, together with those for the Si {\sc iv}  lines at 4089 and 4116~\AA~ in the hotter targets,  and those of {Si} {\sc ii} at 4128, 4130~\AA~ in the cooler stars. For narrow lined, high signal-to-noise ratio targets, a simple Gaussian profile fit was sufficient to obtain a reliable equivalent width measurement (see Fig \ref{fits}, upper left plot). The uncertainties in these measurements are typically of the order of 10\% \citep{hun07}. For stars with higher projected rotational velocities (\vsini\, $\ge $50\kms), it was more appropriate to fit rotationally broadened profiles, as rotation becomes the dominant broadening mechanism (see Fig \ref{fits}, upper right plot). In some cases it was not possible to measure the strength of either the {Si}~{\sc ii} or {Si}~{\sc iv} spectrum. For these targets, upper limits were set on their equivalent widths, allowing constraints to the effective temperatures to be estimated \citep[see, for example,][ for more details]{hun07}. This was mostly the case for stars with large projected rotational velocities and mid-range temperatures (18,000 -- 26,000K).
The random uncertainty in our effective temperatures are approximately $\pm$1000 K (about 5\%), consistent with the high quality of the observational data. In those cases where upper limits have been set for the equivalent widths of the Si absorption lines, the values will obviously be more uncertain and so error bars of $\approx$ 2000 K are more appropriate.

\subsubsection{Surface Gravity}

The logarithmic surface gravity (\logg) was estimated by comparing theoretical and observed profiles of the hydrogen Balmer line H$\delta$. Automated procedures were developed to fit the theoretical spectra to the observed lines, with regions of best fit defined using contour maps of \logg against \teff. Using the effective temperatures deduced by the methods outlined above, the gravity could be estimated. The effects of instrumental, rotational and macroturbulent broadening, which have most significant effect on the line cores,  were considered in the theoretical profiles (see Fig \ref{fits}, lower plots). Uncertainties in the fitting procedures led to random errors of $\pm$ 0.1 dex, while systematic errors could be present due to, for example, the uncertainty in the adopted line broadening theory or in the model atmosphere assumptions. Additional errors may arise from the uncertainty of the identification of the continuum around H$\delta$ in the McDonald spectra (see section \ref{McD_Obs}), where less continuum was available around the line. 

\subsubsection{Microturbulence}\label{vt_meth}

Following standard practice, we derived the microturbulent velocity from the Si {\sc iii} triplet (4552, 4567 and 4574~\AA\ - see upper plots in Fig \ref{fits}) as it is observed in all our analysable spectra and because all three lines arise from the same multiplet, so that errors in the absolute oscillator strengths and non-LTE effects should be minimized. 


\begin{table*} 
\caption{Final atmospheric parameters for each star, listed by HIP number, with another identifier shown. Effective temperatures, surface gravities and microturbulences are given for each star, along with magnesium, silicon, nitrogen and carbon abundances where available$^a$. } 
\label{stars_atmos}
\centering
\begin{tabular}{ccccccccc}   \hline
 \multicolumn{1}{c}{HIP} & \multicolumn{1}{c}{Other} & \multicolumn{1}{c}{$ T_{\rm eff}$} & \multicolumn{1}{c}{$\log g $} & 
 \multicolumn{1}{c}{\vt} & \multicolumn{1}{c}{Mg} & \multicolumn{1}{c}{Si} & \multicolumn{1}{c}{N} & \multicolumn{1}{c}{C}   \\ 
 \multicolumn{1}{c}{} & \multicolumn{1}{c}{} & \multicolumn{1}{c}{K} & \multicolumn{1}{c}{cm-s$^{-2}$} & \multicolumn{1}{c}{km-s$^{-1}$} &
  \multicolumn{1}{c}{} & \multicolumn{1}{c}{} & \multicolumn{1}{c}{} & \multicolumn{1}{c}{}  \\  \hline
      2702 &   HD 3175 &    16100 & 3.6 &  8 & 6.85 & 7.02 & 7.35 & 7.82 \\
      3812 &CD -56 152 &    17000 & 3.4 & 14 & 6.83 & 6.83 & 7.42 & $\ldots$ \\
      7873 &  HD 10747 &    18700 & 3.8 &  8 & 7.03 & 7.01 & 7.33 & 7.69  \\
     13489 &  HD 18100 &    23500 & 3.6 & 15 & 7.05 & 6.64 & 7.42 & $\ldots$ \\
     16758 &  HD 22586 &    21700 & 3.3 & 14 & 7.44 & 7.62 & 7.93 & 8.13  \\
     45563 &  HD 78584 &    18600 & 3.8 &  8 & 7.01 & 6.81 & 7.31 & $\ldots$ \\
     55051 &  HD 97991 &    21500 & 3.8 &  5 & 6.98 & 7.36 & 7.58 & $\ldots$ \\
     56322 & HD 100340 &    24500 & 3.8 &  4 & 7.43 & 7.55 & 7.61 & $\ldots$ \\
     60615 &BD +36 2268&    19600 & 3.4 &  9 & 6.97 & 6.82 & 7.66 & $\ldots$ \\
     61431 & HD 109399 &    23000 & 3.1 & 16 & 7.38 & 7.24 & 7.42 & $\ldots$ \\
     64458 & HD 114569 &    18300 & 3.8 &  6 & 7.21 & 7.24 & 7.90 & 8.39  \\
     67060 & HD 119608 &    19900 & 2.7 & 16 & 7.52 & 7.67 & 7.93 &  8.05 \\
     68297 & HD 121968 &    20550 & 3.4 &  0 & 7.43 & 7.98 & 7.72 & $\ldots$ \\
     70205 &LP  857-24 &    24600 & 4.1 &  0 & 7.45 & 7.46 & 7.42  & $\ldots$ \\
     70275 & HD 125924 &    21000 & 3.6 &  6 & 7.08 & 7.13 & 7.28  & $\ldots$ \\
     79649 & HD 146813 &    19400 & 3.2 &  5 & 7.02 & 7.25 & 7.47  & $\ldots$ \\
     81153 & HD 149363 &    27800 & 3.5 & 12 & 7.60 & 7.79 & 7.86  & 8.22 \\
     85729 & HD 158243 &    19300 & 2.7 & 20 & 7.37 & 7.69 & 7.91 &  8.02\\
     91049 & HD 171871 &    20300 & 3.4 & 14 & 7.35 & 7.51 & 7.81  & $\ldots$ \\
     92152 & HD 173502 &    25600 & 3.5 & 12 & 7.51 & 7.74 & 7.98 & $\ldots$ \\
 
%
 
     94407 & HD 179407 &    26000 & 3.4 & 17 & 8.00 & 7.82 & 8.21 & 8.53 \\
     96130 & HD 183899 &    20000 & 3.3 & 17 & 7.18 & 7.53 & 7.80 & 7.98 \\
     98136 & HD 188618 &    21300 & 3.4 & 11 & 7.32 & 7.34 & 7.85 & $\ldots$ \\
    101328 & HD 195455 &    20550 & 3.2 & 14 & 7.47 & 7.74 & 7.96 & 7.83 \\
    105912 & HD 204076 &    20100 & 3.4 & 20 & 7.19 & 7.43 & 7.97 & - \\
    107027 & HD 206144 &    17750 & 2.5 & 16 & 7.24 & 7.31 & 7.56 &  7.68\\
    109051 & HD 209684 &    20340 & 3.9 &  9 & 6.98 & 7.17 & 7.54 & $\ldots$ \\
    111563 & HD 214080 &    19400 & 2.9 & 17 & 7.18 & 7.63 & 7.56 & 7.81 \\
    112022 & HD 214930 &    18000 & 3.4 &  6 & 7.04 & 7.04 & 7.27  & $\ldots$ \\
    112482 & HD 215733 &    23100 & 2.9 & 14 & 7.39 & 7.40 & 7.60  & $\ldots$ \\
    113735 & HD 217505 &    21600 & 3.9 &  3 & 7.27 & 7.33 & 7.56 & 7.97 \\
    114690 & HD 219188 &    23200 & 3.0 & 13 & 7.37 & 7.71 & 7.66  & 7.86\\
    115347 & HD 220172 &    21700 & 3.8 &  0 & 7.31 & 7.41 & 7.69  & 7.95 \\
    115729 & HD 220787 &    18600 & 3.6 &  5 & 7.01 & 7.07 & 7.51  & 7.91\\
  $\ldots$ &EC 05582-5816 & 15900 & 3.4 & 10 & 6.97 & 7.43 & 8.40 &  8.39 \\
  $\ldots$ &EC 13139-1851&  18100 & 3.9 & 13 & 7.09 & 7.21 & 7.64 & $\ldots$ \\
  $\ldots$ &EC 20140-6935&  21900 & 3.8 &  0 & 7.20 & 7.47 & 7.72 & 8.09  \\
  $\ldots$ &   PB 5418 &    19300 & 3.8 &  8 & 7.05 & 7.07 & 7.44  & $\ldots$ \\
  $\ldots$ &   PHL 159 &    22900 & 4.1 &  0 & 7.46 & 7.41 & 7.46 & $\ldots$ \\

\hline

\end{tabular}
\flushleft { $^a$ Analysis of EC 05582-5816 is based on model atmospheres with the Si abundance set to 7.4. For all other
analyses, the Si abundance is determined from silicon lines and the microturbulence from the Si\,{\sc iii} lines. }
 
 \end{table*}

An alternative approach to the microturbulence determination  uses the same Si\,{\sc iii} triplet but  finds the microturbulence that gives the baseline Si abundance of 7.42 from \citet{hun09a} as the average from the three lines.  This alternative impacts also the determination of
the effective temperature and surface gravity and, hence, the abundances of all elements.  In the next section, we comment on the
effect on the abundances of the two methods for determining the microturbulence.



\subsection{Elemental abundances}

Nitrogen, magnesium, silicon, and in some cases carbon abundances for each star have been estimated using the atmospheric parameters given in Table \ref{stars_atmos} and  measurements of absorption lines. Abundances are given in Table \ref{stars_atmos}.
The atmospheric parameters agree well between observations with the different spectrographs. Effective temperatures show differences not larger than 1200 K in all cases, while the range of the logarithmic gravity estimates is only greater than 0.2 dex in one instance (HD 121968 with a range of 0.25 dex). Differences in microturbulence are less than 6\,\kms\ in all cases. The agreement between atmospheric parameters derived from observations obtained with different instruments is reassuring but not surprising given the high-quality of all spectra. This agreement is, of course, not a measure of any systematic errors.
 
Extensive appendices in \cite{hun07} show how errors in the atmospheric parameters for B-type  stars translate into errors in derived abundances. Hunter et al. considered the errors to be independent but in reality the situation will be more complicated, as this is not the case. For example, an increase in the effective temperature estimate will lead to an increase in the gravity estimate and this leads to the theoretical N II equivalent widths (and hence nitrogen abundances) being less sensitive to changes in the atmospheric parameters than if these are considered to vary independently.

For the range in atmospheric parameters found for our sample, the simulations of Hunter et al. imply (see for example their Fig. 6 for the N II 3995~\AA\, line) that our estimated errors in both the gravity and microturbulence lead to relatively small errors in the nitrogen abundance estimate of 0.1 dex or less. Errors in effective temperature estimates are more important and lead to larger uncertainties with an error of $\Delta$\teff$\pm$~1000K translating into an nitrogen abundance error of approximately $ \mp$ 0.2 dex at an effective temperature of 18000 K but decreasing to $\sim$0.1 dex at an effective temperature of 25000 K. Additionally there will be random errors in the nitrogen abundance estimates due to uncertainties in the N II equivalent widths. The latter have been estimated as $\pm$10\% which would imply an uncertainty of $\sim$ 0.1 dex. 
 Combining these different sources of error in quadrature would lead to typical uncertainties of 0.2-0.3 dex.

For the atmospheric parameters and abundances in Table \ref{stars_atmos}, the microturbulence was derived from  the three Si\,{\sc iii} lines with the condition that the three lines return the same Si$^{2+}$ abundance. This method was successful for all but  EC 05582-5816 for which lines are greatly rotationally broadened. As we noted above, the alternative method
is to assume a standard Si abundance of 7.4 for the  
 determination of the microturbulence and, hence, other atmospheric parameters which finally define the model atmosphere used to determine the abundances of C, N and Mg. (This alternative method
was used by \cite{hun07} in their analysis of B stars in three Galactic clusters.) The C, N and Mg abundances are little affected by how the microturbulence
is chosen; the preferred and alternative methods give similar results. This is well shown by Figure 2  where we show the N abundances obtained when the microturbulence is set by the condition
that the Si abundance is 7.4 plotted versus the abundances obtained when the microturbulence comes from  the
three Si\,{\sc iii} lines. It is seen that the N  abundances are not systematically different
in the two cases except for two or three  outliers. Obvious outliers are HD 18100 and BD +36 2268 with N abundances from the alternative method of
7.97 and 8.20, respectively.  A similar correspondence between abundances from the two methods is found for Mg and the limited number or determinations of
the C abundances.  Thus, apart from the artificial constraint of a constant Si abundance, the abundance analyses for all other elements are
expected to be largely unaffected by the method used to determine the microturbulence and, hence, the atmospheric parameters.




For a subsample of our stars (all stars observed with FEROS) we estimated carbon abundances.  
 The C {\sc ii} line at 4267\,\AA\, was used to calculate abundances as it is the only measureable line in all of the spectra. This line is known to be susceptible to (subtle) non-LTE effects \citep{nie06, nie08}, which are not fully taken into account in the model ion that was included in the {\sc TLUSTY} calculations - see discussion  by \cite{hun09a}. Our interpretation (see below) of the various abundances is in part referenced to the abundance analysis of Galactic
cluster by \cite{hun07} who used also exclusively the 4267\,\AA\ line and, thus, the inability  to account fully for the non-LTE effects should be almost
cancelled by the comparison with the Galactic clusters.

\begin{figure}
\begin{center}
 \includegraphics[trim=0.4cm 0.2cm 0.9cm 0.8cm, clip=true,height=0.32\textheight,width=0.48\textwidth]{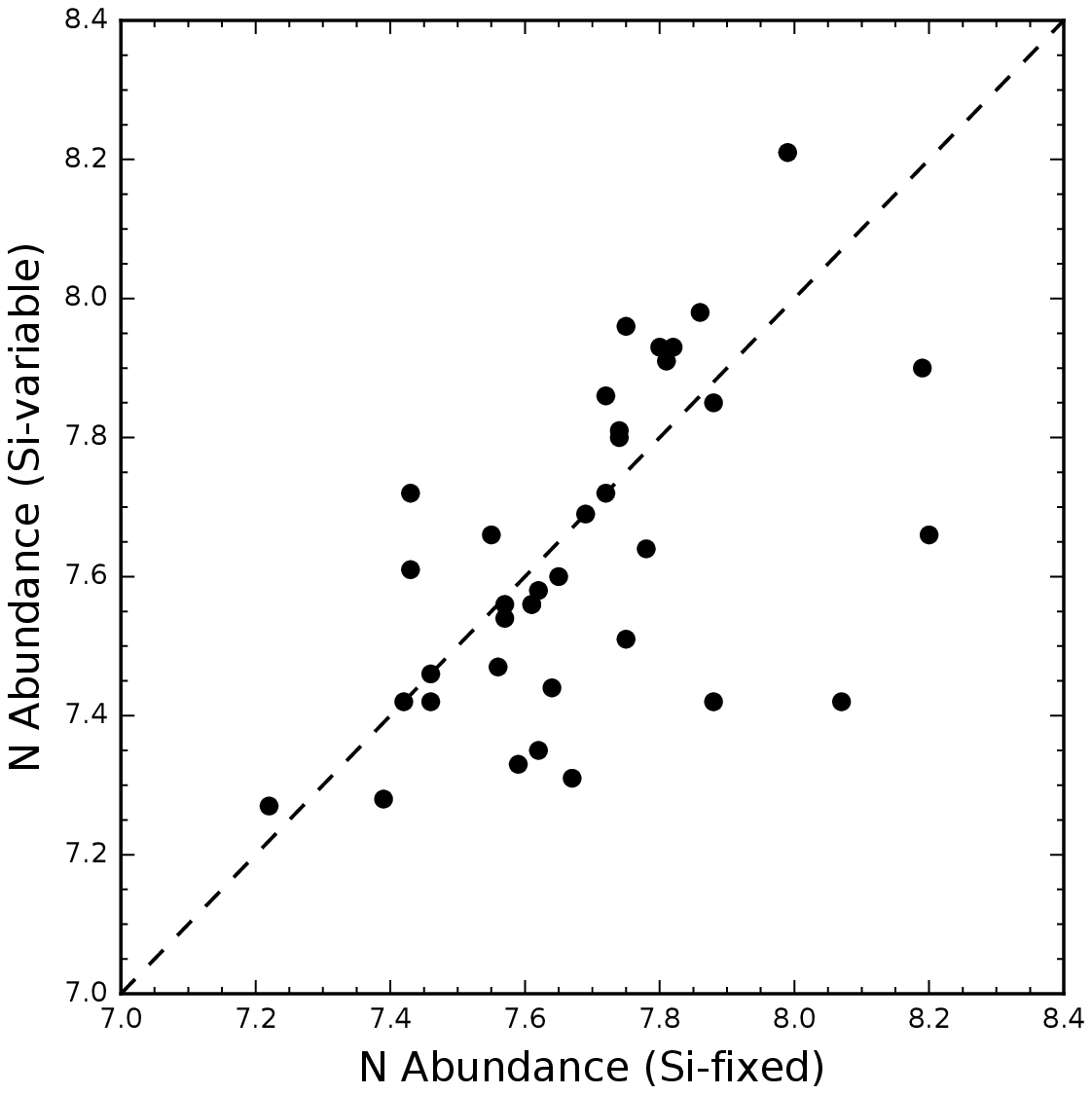}
\caption{Nitrogen abundances from models with the microturbulence set by the condition that the Si abundance is 7.40 versus the N abundances set by determination of the microturbulence from the Si\,{\sc iii} lines. The solid line of unit slope corresponds to an abundance independent of the method of fixing the microturbulence.  }
 \label{NMg}
%
 \includegraphics[trim=1.6cm 0.2cm 0.2cm 1.5cm, clip=true,height=0.30\textheight,width=0.49\textwidth]{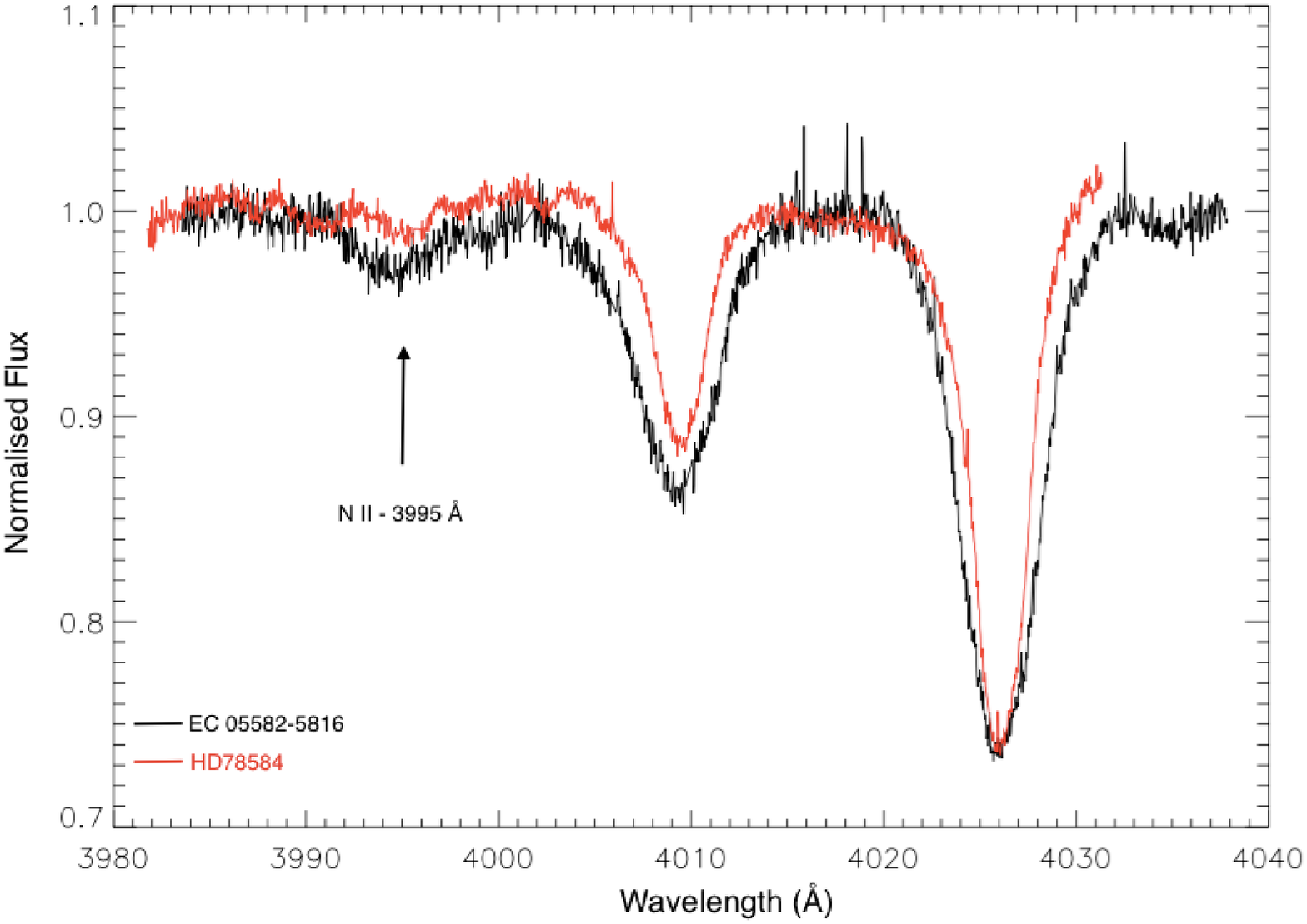}
\caption{The spectrum around the N {\sc ii} line at 3995\,\AA\, for the most nitrogen enhanced star, EC 05582-5816. HD 78584 is over-plotted in red for comparison, as it has similar spectral type (B3) and \vsini\, of 102\,\kms. Although the nitrogen line is heavily broadened in EC 05582-5816, due to its very high \vsini\, of 221\,\kms, it can still clearly be seen, so has to be particularly strong to avoid being smeared out into the surrounding continuum.} 
 \label{f_N}
\end{center}
\end{figure}


Nitrogen abundances in Table \ref{stars_atmos} were estimated  from the singlet transition at 3995~\AA~ as it is one of the strongest N {\sc ii} lines in the optical spectrum and appears unblended.
Two examples of  spectra of high \vsini\, stars around the N line at 3995~\AA\, are shown in Figure \ref{f_N}.
Other N {\sc ii} lines were also present in a significant number of observations, including those between 4601 and 4643\,\AA~ and the singlet at 4447~\AA. These tended to be more blended, and so may lead to less accurate abundance estimates than those from 3995~\AA. 
 It is interesting to note that \cite{lyu13} found that the line at 3995~\AA\, gave a lower N abundance in their sample compared with the abundance from other transitions and that the differences were a function of effective temperature ranging from $-0.3$ dex at 16000 K to 0.0 dex at 29000 K. In our case, the difference is about $-0.1$ dex over the full temperature range with  a rise to 0.0 dex below about 18000 K.  These differences with \cite{lyu13} likely reflect differences in the adopted model atoms. By referencing our abundances from the 3995~\AA\ line to those from \cite{hun09a}, we expect to obtain a true measure of abundance differences between the cluster (i.e., Galactic disk)  and runaway B stars.


Magnesium abundances were  estimated for all spectra using the Mg {\sc ii} transition at 4481\AA\ comprising three overlapping lines of a single multiplet. 

Silicon abundances are obtained from the lines of Si\,{\sc ii}, Si\,{\sc iii} and Si\,{\sc iv} referred to in the determination of the atmospheric parameters.

\section{Ejection mechanisms and abundances}

The most satisfying discussion of the abundances would conclude with the
demonstration that runaway B stars formed by the BSS and the CES
have distinct differences in composition with both scenarios providing
abundances differences with B stars in Galactic open clusters.
Perhaps the least satisfying result would be one in which the runaway B
stars showed a common pattern of abundances and one which matched
well the abundances of the B stars in Galactic open clusters, i.e., the BSS and CES scenarios both preserve the abundances of the investigated elements. Yet in this case,
 the origin of runaway stars may hopefully be traced from
other observational indicators such as binarity and rotational velocities.

The X-ray sources known as low-mass X-ray binaries (LMXBs)  consist of a low mass ``normal'' secondary star orbiting a black hole (BH) or a
neutron star (NS) where the supernova leading to the BH or NS likely contaminated the secondary.  Abundance analysis of the secondary
should offer clues to the contamination expected in the BSS mode of runaway star formation.
One LMXB secondary accompanying a BH has a spectral type of B9III and with an effective temperature of 10500 K approximates a runaway star.  With [Fe/H] = 0.0, this secondary  in V4641 Sgr has a normal composition (C, O, Mg, Al, Si and Ti)  but an +0.8 dex overabundance for N and Na \citep{sadakane06}. Curiously, this star appears to be a replica of EC 05582-5816 (see Section 5.3). Unfortunately C and N abundances have not been reported for the other six LMXB secondaries. The secondary with the most extreme enrichments would appear to be Nova Sco 1994 \citep{gonzalez08} with, for example, [Fe/H] $= -0.1$ but  [Mg/Fe] $= +0.4$ and [Si/Fe] $= +0.7$ and other anomalies including [O/Fe] $= +1.0$ and [S/Fe] $= +0.9$. Mg and Si abundance estimates are available for five secondaries.  Comparisons with normal F, G and K dwarfs are made in the [Fe/H] versus [X/Fe] plane where X = Mg or Si. The  mean [Fe/H] for the five stars is $+0.1$ with a spread from $-0.1$ to $+0.3$. The mean difference with respect to normal stars of the same [Fe/H] is [Mg/Fe] and [Si/Fe] at $+0.2$ with a spread from 0.0 to about $+0.7$ but the Mg and Si enrichments would be larger if Fe were also enriched.  There is a hint that [Mg/Fe] and [Si/Fe] are positively correlated.  Other elements considered for four or more of the secondaries but not included in our
analysis include the following with  mean  differences of [X/Fe] relative to normal stars: $+0.4$ (O), $+0.4$ (Na), $+0.1$ (Al), 0.0 (Ca), $+0.2$ (Ti) and $+0.1$ (Ni), again elemental abundance enrichments are larger if supernova Fe contaminates the secondary.  Predicted
enrichments depend on many factors including uncertain aspects of the SN explosion and do not completely account for the
composition of these secondaries.   In summary, as far as the elements considered here are concerned,  observations of C and N are
too few to suggest a pattern but Mg and Si may be enriched simultaneously in some cases.

Possible changes of composition occurring at the birth of a runaway star
have to be extracted from the observed composition bearing in mind that two
other factors may affect the chemical composition of the B stars prior to their conversion to a
runaway star:
\noindent First, the initial composition of B (and other) stars depends on the location
of their birth site; studies of the composition of stars and H\,{\sc ii}
regions show that abundances decline with increasing distance from the
Galactic center. For example, \cite{daf04} from B stars and
\citet{luck2011} from Cepheids find that the abundance gradients for
our elements are about $-0.045$ dex kpc$^{-1}$, a value representative of
many other studies. Azimuthal variations are considered to be smaller than
the radial variation.  Since runaway stars in the sample catalogs are expected to have
birth sites spanning several kpc in the Galactic disk, abundance variations of several 0.1 dex may be
anticipated. 
\noindent  Second, B stars may be prone to
mixing between the surface and the interior resulting in changes of
composition as CN-cycled H-burning products reach the atmosphere, i.e.,
the He/H ratio is increased with a coupled decrease of C and an increase of
the N abundance -see, for example, calculations of surface abundances along evolutionary 
tracks for rotating massive stars by \cite{brott11}. 
(Extreme mixing may contaminate the atmosphere with ON-cycled H-burning
products resulting in decreases of O and additional increases of N and He.)

\subsection{Magnesium and Silicon}

Since the Mg and Si abundance should be unaffected by mixing within
evolving B stars but could conceivably
be altered in runaway stars created by the
BSS, we discuss this pair of elements ahead of our discussion of C and N. Our
reference Galactic B stars are the three open clusters analyzed by
\cite{hun09a}: NGC 6611 at a Galactocentric distance $R_G$ of 6.1 kpc,
NGC 3293 at $R_G = 7.6$ kpc and NGC 4755 at $R_G = 8.2$ kpc where
$R_G$ is taken from Rolleston et al. (2000). (Daflon \& Cunha (2004)
give somewhat different estimates.)  Abundances provided by \cite{hun09a} correspond to
atmospheric parameters chosen by assuming a Si abundance of 7.4 (see above for comments on such a choice).
Since the 2.1 kpc difference in Galactocentric distance across the three open clusters should correspond to an abundance difference of 
only about 0.09 dex, assumption of constant Si abundance is likely not a source of significant systematic error. 
The Mg abundances for the three Galactic clusters are just consistent with the
anticipated 0.09 dex decline between NGC 6611 and NGC 4755. Of course, the Si abundances are not expected to betray the
abundance gradient because the condition Si = 7.4 was imposed on the analysis.



For the runaway B stars, the Mg and Si abundances from Table \ref{stars_atmos} are compared in Figure 4 which shows that Mg and Si abundances are
highly correlated and each range over 1.0 dex. The straight line in Figure 4 corresponds to a constant Mg/Si ratio set at the `standard' ratio of
$-0.17$ dex adopted by \cite{hun09a}. The spread in Mg abundances among the runaway stars far exceeds the range
among stars from an individual open cluster which is a fair measure of the measurement uncertainties given that stars within a cluster share a common Mg abundance.
A major  contribution to the Mg and Si abundance spread among the runaway stars
surely  arises because their birthplaces  span quite a range in
Galactocentric distances even though present Galactocentric distances may not be too different, a range  much greater than the range spanned by the three reference open clusters.
 Indeed, Galactic orbits calculated by Silva \& Napiwotzki
(2011) give birthplaces from the Galactic center out to nearly 14 kpc for the sample drawn from their paper. However, plots of the Mg and Si abundances versus the estimated Galactocentric distance of a star's birthplace are
too ill-defined to confirm the abundance gradients obtained from {\it in situ} abundance measurements of young stars and
H\,{\sc ii} regions. All but three of our stars have estimated birthplaces between 4 kpc and 10 kpc.
The failure to reproduce the observed abundance gradients may be attributable to uncertainties in the locations of the birthplaces  and to
contamination of the stars in the BSS  by Mg and Si from the supernova. 

Scatter in Figure 4 about the linear relation is consistent with Mg and Si abundance uncertainties and with the
spread in Mg abundances within each of the open clusters (see below). Presence of the linear correlation
and a constant Mg/Si ratio is qualitatively expected from the similar nucleosynthetic yields of Mg and Si  from Type II supernovae which
are surely the controlling influence on the Galactic abundance gradient whatever the influence of the myriad other factors  (e.g., the initial
mass function, the star formation rate,  Galactic infall etc.) entering recipes for Galactic chemical evolution.


\begin{figure}
\begin{center}
 \includegraphics[trim=0.4cm 0.2cm 0.9cm 0.8cm, clip=true,height=0.32\textheight,width=0.48\textwidth]{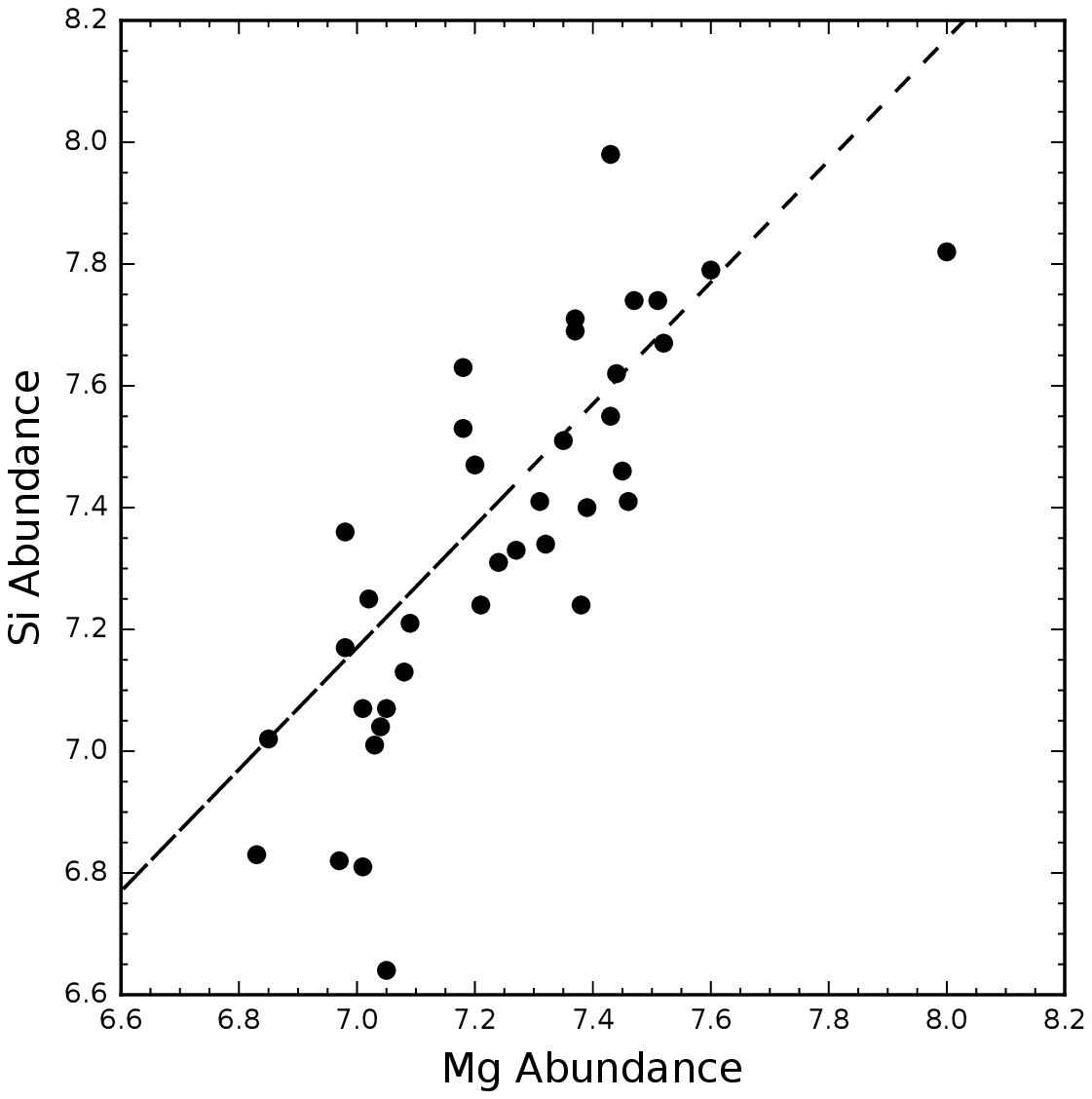}
 \caption{Mg and Si abundances from Table \ref{stars_atmos}. The straight line corresponds to a Mg/Si ratio of $-0.17$ dex normalized to the standard abundances adopted by \cite{hun09a}.}
 \label{f_MgSi}
\vspace{0.1cm} 
%
%
 \includegraphics[trim=0.4cm 0.2cm 0.9cm 1.2cm, clip=true,height=0.30\textheight,width=0.49\textwidth]{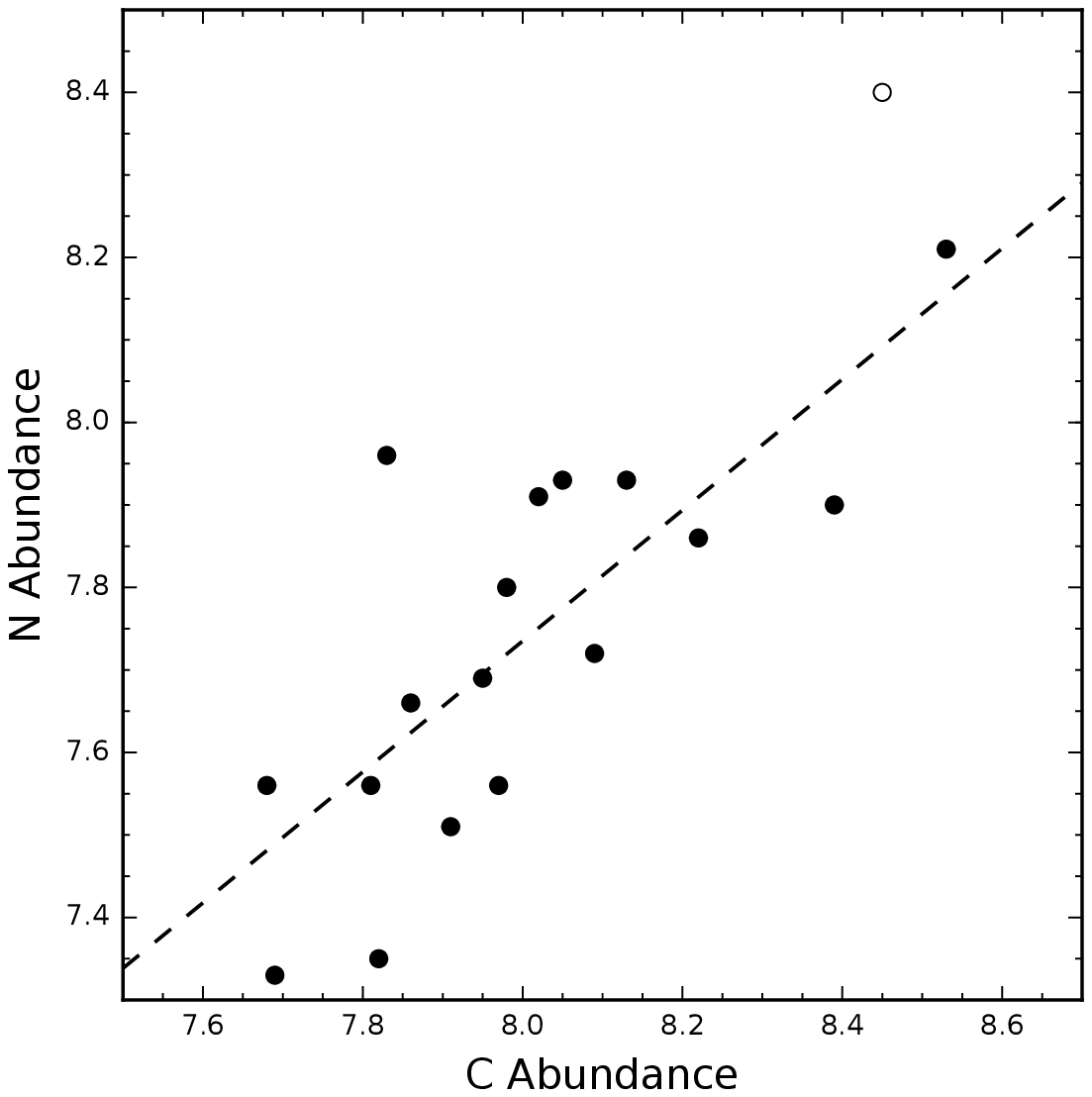}
 \caption{C and N abundances plotted for the {\it FEROS} sample of runaways. Abundances are taken from Table \ref{stars_atmos}. The straight line, a least-squares fit,  e shows
that C and N abundances are  related over a range of about 1 dex with a scatter consistent with measurement uncertainties, i.e., there are
no stars heavily contaminated with CN-cycled products, that is stars which are C-poor and N-rich. Data for EC 05582-5816  (open circle) are taken from the analysis in which the Si abundance is fixed at 7.4.  }
 \label{carbon}
\end{center}
\end{figure}

\subsection{Carbon and Nitrogen}

Among the B stars in the three reference clusters, \cite{hun09a} found that
the C and N abundances were  independent of
position on a star's evolutionary track except that half of the few
supergiants were enriched in CN-cycled material. Similarly, \cite{lyu13}
found normal C and N abundances in a sample of 22 slightly-evolved
B stars (i.e., no supergiants)  except for two
possibly mixed stars. Thus, the  expectation is that the majority of the  B stars
before undergoing conversion  to a runaway star would have
preserved their initial C and N abundances. The likely corollary of this expectation is that alterations to the C and N abundances
may be clues to the process creating the runaway star but internal CN-cycling and mixing to the surface may have occurred independently of formation as a runaway star.

As noted above, C abundances in Table \ref{stars_atmos} are provided only for the stars observed with FEROS. 
Figure 5 shows the C and N abundances for this minority.   C and N abundances for EC 05582-5816 in Table \ref{stars_atmos} unlike for other stars 
are  taken from an analysis with
the Si abundance fixed at 7.4 but,  as noted above,  the C and N abundances are expected to be independent of this assumption
concerning the Si abundance; the high $v\sin i$ of EC 05582-5816 precluded the use of the Si\,{\sc iii} multiplet to determine the
microturbulence. The C and N abundances are tightly correlated with all stars showing a similar  C/N ratio to within the measurement
uncertainties. In addition, the three Galactic clusters analysed by \cite{hun09a} have mean C and N abundances falling within the
band defined by the runaway stars, that is mean abundances of 8.00$\pm0.19$ and 7.62$\pm0.12$ for C and N, respectively, for all non-supergiants
in the three Galactic clusters with these abundances provided by analyses assuming Si has its standard abundance.


These carbon abundances for the subset of our runaway stars show no evidence for a runaway star enriched in CN-cycled material, i.e., a low C
abundance paired with a high N abundance such that the total number of C and N atoms is conserved. Very severe contamination with
CN-cycled material would result in a N enrichment of up to 0.5-0.6 dex and a severe depletion of C. No such stars are seen in Figure 5. Presence of CN-cycled
material may occur in the B star prior to its conversion to a runaway star and is, thus, not a determining signature of a process resulting in
a runaway star. 

\subsection{Nitrogen and magnesium}

 To extend the search for abundances anomalies to the complete sample, we
examine next the relationship between the N and the Mg abundances. 
Figure 6 drawing on Table 3 compares the N and Mg abundances with a
distinction made according to surface gravity (i.e., evolutionary status)
with stars with $\log$ g $<$ 3.2 (i.e., supergiants) represented by open circles
and stars of higher surface gravity (i.e., main sequence and slightly 
evolved stars) represented by filled circles. Figure 7 shows the N and Mg
abundances for B stars belonging to the cluster NGC 3293, the best represented
of the three clusters studied by Hunter et al. (2009). 

With the possible exception of N enrichment from internal mixing or mass loss,
one expects the cluster's stars in Figure 7 to share the same N and certainly
the same Mg abundance. Thus, the scatter in N and Mg abundances
in Figure 7 represents the
measurement uncertainties  which, thanks to the similarity of analytical
techniques, will be a very close approximation to the uncertainties affecting
the points in Figure 6. (One star -- a supergiant --  in NGC 3293 
 appears N-enriched but this star is not C-depleted.) 
Comparison of Figures 6 and 7 shows that the
spread shown by NGC 3293's B stars and, hence, we conclude that the runaways
show an intrinsic star-to-star difference in composition (i.e., C, N, Mg and Si
abundances but with similar abundance ratios) which we attribute to differences
in a star's birthplace with the Galactic abundance gradient being a very
likely controlling factor. 

EC 05582-5816 and HD 179407 appear as outliers in Figure 6.  The two outliers have similar compositions but for their Mg
abundances and both  appear related to the inner Galaxy. Silva \& Napiwotzi (2011) estimate EC 05582-5816's
birthplace at 2 pc from the Galactic centre. HD 179407 is presently at $R_G = 3.5$ kpc according to \cite{smartt1997}.  
The pair have very similar C and N abundances (Table 3)  but differ substantially in Mg with an abundance of 8.0 for
HD 179407 but 7.0 for EC 05582-5816. The Si abundance of HD 179407 is consistent with its Mg abundance.
For EC 05582-5816, present analysis assumes a Si abundance of 7.4. Galactic abundance gradients likely account for
the high abundances found for HD 179407.   There is a remarkable similarity in composition between EC 05582-5816 and the
B9 III secondary of the black hole binary V4641 Sgr.  This B star has [Fe/H] = 0 and normal abundances of C, O Mg, Al Si and Ti but a high N
and Na overabundances ([N/H] = [Na/H] = +0.8) \citep{sadakane06}.  If placed in Figure 6, it would provide a third outlier and fall
near the two existing outliers! Reobservation and  reanalysis of EC 05582-5816, a rapid rotator is to be
encouraged and extended to additional elements.

\begin{figure}
\begin{center}
 \includegraphics[trim=0.5cm 7.5cm 4.9cm 4.0cm, clip=true,height=0.30\textheight,width=0.8\textwidth]{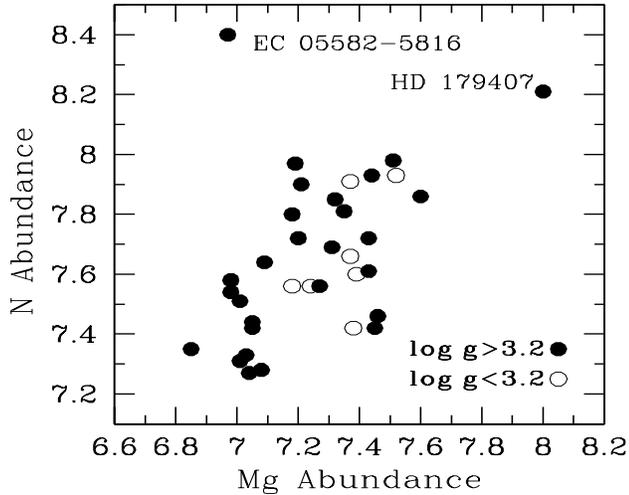}
\caption{Mg and N abundances from Table 3 with stars distinguished by their surface gravity: $\log$ g $> 3.2$ shown as filled circles and $\log$ g $< 3.2$ shown as unfilled circles. Two outliers are identified $--$ EC 05582-5816 and HD 179407 $--$ and discussed in the text. } 
 \label{n-mg}
\end{center}
\end{figure}

\begin{figure}
\begin{center}
 \includegraphics[trim=0.5cm 7.5cm 4.9cm 4.0cm, clip=true,height=0.30\textheight,width=0.8\textwidth]{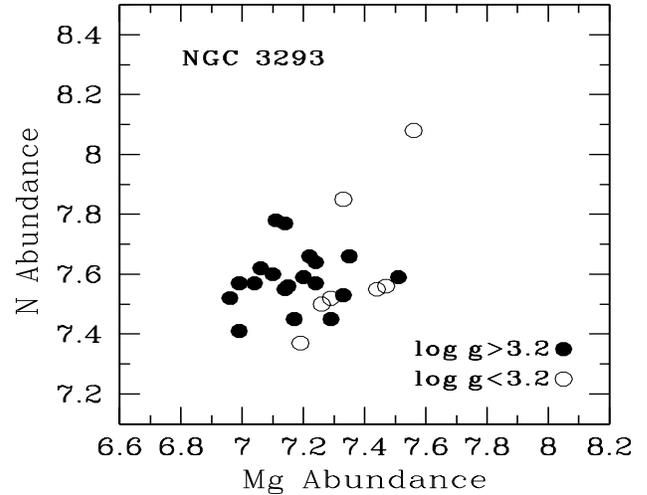}
\caption{Mg and N abundances for B stars in the open cluster NGC 3293 with data from Hunter et al. (2009). Stars are distinguished by their surface gravity $--$ see caption to Figure 6. } 
 \label{n-mg-3293}
\end{center}
\end{figure}

\subsection{Formation mechanisms and Abundance Anomalies?}

Nitrogen, magnesium and silicon abundances across the sampled runaway B stars
are similar to those seen in B stars in Galactic clusters. The spread in the C, N, Mg and Si abundances and the nearly uniform abundance
ratios C/N/Mg/Si  among runaway stars in our sample results from the combination of two facts:   (i) the stars' birthplaces sample a wide range in
Galactocentric distances  and (ii)  the abundances in star-forming regions decline with increasing Galactocentric
distance  at a rate of approximately 0.05 dex kpc$^{-1}$. Within our sample, there are no certain outliers with abundance
peculiarities. 
Thus,  the conclusion is  that the two scenarios -- BSS and CES -- capable of
 ejecting  B (and other) stars from the Galactic disk into the Galactic halo  do not as a rule
change the surface chemical
composition of the runaway star, as sampled by C, N, Mg and Si.  This conclusion from our non-LTE analysis confirms earlier
LTE analyses in which the compositions of a runaway star and a comparable B star in the Galactic disk are compared, see, for
example, \citet{martin2004}.  Although changes may occur in rare cases,
the general conclusion is both disappointing and challenging. Disappointing in that
the composition, a readily obtainable quantity,  appears not to be a discriminant
between competing ejection mechanisms. Challenging in that other observable
quantities now have to  been relied upon to identify the principal  ejection mechanism across a sample of runaway stars and for individual
runaway stars.  Possible observables include projected rotational
velocities and binarity.

\section{Ejection mechanisms:  rotational and radial velocities}

\subsection{Projected rotational velocities}

The distribution of projected rotational velocities $v\sin i$ for our sample of runaways is
very similar to that provided by the B stars in the three Galactic clusters observed by
\cite{hun09a}: velocities up to 300 km s$^{-1}$  with a peak near 50 km s$^{-1}$ and few slow
rotators (say $v\sin i \leq 30$ km s$^{-1}$. \citet{martin2006} considered rotational velocities
for his sample of runaways and found their frequency distribution to be similar to that assembled
by \citet{guthrie1984} from ``young'' OB associations whose result is similar to that from the
three Galactic clusters. In contrast, B stars in ``old'' associations and field B stars in the Galactic disk  have  distributions  which rise with
decreasing $v\sin i$ \citep{wolff1982,guthrie1984}. A more recent catalog of projected rotational velocities for 102
northern B stars is provided by \cite{abt02}.

Tying the $v\sin i$ distribution of the runaways to the ejection
mechanisms is unfortunately hampered by the absence of quantitative predictions for the BSS and CES.
In the BSS, if the runaway originates in a close binary system, the runaway may be tidally locked
resulting in the rotation velocity being closely related to its orbital velocity. In turn, the ejection
velocity is expected to be  related to the orbital velocity. These ideas, lead to the
expectation that the ejection velocity should be positively correlated with the rotational velocity but, as
\citet{martin2006} demonstrated runaway stars do not show the expected correlation and, therefore, the BSS is not the
leading producer of runaway stars.
For the CES, the likelihood of ejection is plausibly  higher in denser environments. On this simple premise and the
match between the $v\sin i$ frequency distribution  for runaways and ``young'' associations, Martin  considered
that it `seems likely that the [CES] is the dominant mechanism'' behind his runaway sample and by extension
ours too which has overlap with his sample.  

\subsection{Radial velocities}

In the BSS, the resulting runaway B star will be either a single B star or form a binary system with
 a low-mass compact companion (e.g., a neutron star).  A runaway which is single will
be paired with a distant   ejected neutron star or a black hole moving in a generally opposite direction. 
Portegies Zwart (2000) suggests that 20\% to 40\% of the runaway B stars should be accompanied by
a neutron star and, thus, the runaway B star should appear as a single-lined spectroscopic binary with a period of
several hundred days.

In the CES, simulations  \citep{leodun1990} suggest that 10\% of the runaways will be binaries comprised
of normal main sequence stars and likely observable as double-lined spectroscopic binaries. 

A definitive test of these predictions for the BSS and CES will require both a radial velocity long-term survey of a
large sample of runaway B stars to be put up against more precise predictions about the frequency and nature of the
binary populations from the two scenarios.  Then, it may be possible to assess in a statistical
fashion the relative production rates from the BSS and CES. For the binaries, it should be possible to
assign them to either the BSS (i.e., a neutron star/pulsar companion) or to the CES (i.e., a normal
main sequence companion). For the runaway single stars, the attribution to the formation scenario
will be difficult unless one can uncover a subtle abundance anomaly or apply precise determination of
space motions (GAIA?) to   assist in  the identification of the pulsar or the original stellar association.

A definitive observational test is not possible at present. The sole radial velocity survey 
of high Galactic OB stars was conducted by Gies \& Bolton (1986) who
concluded  that ``runaway OB stars are deficient in close binaries by a factor of 2-4''.  Their sample of
15 confirmed  runaways provided two binaries (both double-lined systems) and, if five probable runways are added, the
binary probability becomes 2/20 or 10\%. The deficiency of binaries is suggested
by reference to the binary fraction  of 31\% among normal O stars \citep{garmany1980} and
of 38\% among normal B stars \citep{abt1978}. Five of our stars (HD 97991, 149363, 214930, and 219188) were in
the Gies \& Bolton survey and declared by them to have a constant radial velocity. \citet{martin2003} found the
runaway HD 138503 to be a double-lined spectroscopic binary, a star in Silva \& Napiwotzki's list of runaway
stars.  HD 1999 and HD 204076 observed by us are double-lined binaries. Two radio surveys for pulsars  reported no detections coincident with runaway OB stars \citep{sayer1996,philp1996}. Unfortunately, the correction for the beaming of pulsar radiation results in an uninteresting  limit on the fraction of runaway stars with low-mass compact companions; Sayer et al. estimate that less than 25-50\% of OB runaways
have a neutron star companion, an estimate consistent with the prediction by Portegies Zwart (2000).

Our sample can make only a modest contribution to the frequency of binary runaway stars as a test of the BSS and CES. For the majority of our sample, the star was observed once and an assessment of the radial velocity variation must come from velocities reported in the literature. Eight stars  were observed at two or more telescopes and, thus, at different epochs and the two or three
radial velocity measurements may be inter-compared and also checked against the literature. (This restricted search for
velocity variations is akin to that reported by Martin (2006) who observed many runaway stars at least twice over a
few days and also checked velocities against the literature.)  We have defined a star with a radial velocity variation between measurements of greater than 20 km s$^{-1}$ as a possible member of a binary system.  This cut off is set by
inspection of other studies of OB stars \citep{sana2013,dunstall2015}. In some stars, significant velocity variations arise from pulsations which complicates the search for orbital radial velocity variations and, if neglected as a source of radial velocity variations, the fraction of spectroscopic binaries is overestimated. 


\begin{table*} 
\caption{Each star listed by HIP number along with alternative identifier, radial velocity from our analysis and from the literature (with references) along with the binary status of the star. } \vspace{0.2cm}
\label{stars_rv}
\centering
\begin{tabular}{cccccc}   \hline
 \multicolumn{1}{c}{HIP} &\multicolumn{1}{c}{Other} & \multicolumn{2}{c}{Radial velocity} &\multicolumn{1}{c}{Ref$^{b}$} &
 \multicolumn{1}{c}{Status}   \\ 
 \multicolumn{1}{c}{} & \multicolumn{1}{c}{} & \multicolumn{1}{c}{Us$^{a}$} & \multicolumn{1}{c}{Literature} & \multicolumn{1}{c}{} & 
 \multicolumn{1}{c}{}    \\ \hline
  2702 & HD 3175      & -13$\pm$2 (F)               &          -16$\pm$3   & R3 &  Single \\  
  3812 & CD -56 152   &  14$\pm$8 (U)               &           19$\pm$10  & R2 &  Single \\ 
  7873 & HD 10747     &  -9$\pm$2 (F)               &          -12$\pm$2   & R1 &  Single \\     
 13489 & HD 18100     &  80$\pm$7 (F)               &           76$\pm$3   & R1 &  Single \\ 
 16758 & HD 22586     &  99$\pm$1 (F)               &           97$\pm$2   & R3 &  Single \\
 45563 & HD 78584     & -120$\pm$6 (T)               &         -125$\pm$2  & R1 &  Single \\
 55051 & HD 97991     &   31$\pm$3 (U)               &            26$\pm$3 & R3 &  Single \\
 56322 & HD 100340    &  253$\pm$10 (T), 263$\pm$4 (U) &          254$\pm$9 & R2 &  Single \\
 60615 & BD +36 2268  &   31$\pm$4 (T)               &            31$\pm$7 & R3 &  Single \\
 61431 & HD 109399    &  -43$\pm$3 (F)               &           -49$\pm$2 & R1 &  Single \\
 64458 & HD 114569    &  104$\pm$2 (F)               &            $\ldots$  &  É & Unknown \\
 67060 & HD 119608    &   31$\pm$1 (F)               &            26$\pm$4  & R1 &  Single \\
 68297 & HD 121968    &   17$\pm$9 (T), 29$\pm$3 (U)  &            28$\pm$2  & R4 &  Single \\
 70205 & LP 857-24    &  243$\pm$4 (F)               &           -54$\pm$2  & R5 &  Binary? \\
 70275 & HD 125924    &  244$\pm$1 (T)               &           239$\pm$2  & R4 &  Single \\ 
 79649 & HD 146813    &   21$\pm$2 (T)               &            19$\pm$6  & R4 &  Single \\
 81153 & HD 149363    &  145$\pm$3 (T), 146$\pm$3 (U), 144$\pm$3 (F) & 141$\pm$2 & R1 & Single \\
 85729 & HD 158243    &  -63$\pm$2 (F)               &           -64$\pm$3  & R1 &  Single \\    
 91049 & HD 171871    &  -64$\pm$1 (T)               &           -62$\pm$5  & R1 &  Single \\
 92152 & HD 173502    &   49$\pm$1 (F)               &            68$\pm$4  & R1 &  Single? \\
%

 94407 & HD 179407    &  -120$\pm$4 (F)              &          -119$\pm$5  & R6 &  Single \\
 96130 & HD 183899    &   -46$\pm$2 (F)              &           -45$\pm$5  & R1 &  Single \\
 98136 & HD 188618    &    46$\pm$4 (F)              &           -15$\pm$5  & R1 &  Binary? $^{c}$ \\
101328 & HD 195455    &    19$\pm$7 (F), 10$\pm$6 (U) &            10$\pm$6  & R3 &  Single \\
105912 & HD 204076    &     0$\pm$3 (F), 14$\pm$2 (U) &           -17$\pm$7  & R2 &  Binary  \\
107027 & HD 206144    &  122$\pm$5 (F), 121$\pm$2 (U) &           117$\pm$8  & R4 &  Single \\
109051 & HD 209684    &   82$\pm$2 (U)               &            72$\pm$8  & R4 &  Single \\
111563 & HD 214080    &   16$\pm$2 (F)               &            12$\pm$4  & R6 &  Single \\
112022 & HD 214930    &  -60$\pm$4 (T)               &           -63$\pm$2  & R3 &  Single \\
112482 & HD 215733    &   -6$\pm$6 (T)               &           -15$\pm$2  & R3 &  Single \\
113735 & HD 217505    &  -17$\pm$1 (F)               &           -31$\pm$10 & R2 &  Single \\
114690 & HD 219188    &   73$\pm$19 (F), 98$\pm$19 (T) &           64$\pm$3  & R1 &  Single? $^{d}$ \\
115347 & HD 220172    &   26$\pm$2 (F)               &            29$\pm$3  & R6 &  Single \\
115729 & HD 220787    &   26$\pm$2 (F)               &            26$\pm$3  & R4 &  Single $^{e}$ \\
$\ldots$ & EC 05582-5816&   85$\pm$13 (F)             &            81$\pm$10 & R2 &  Single \\
$\ldots$ & EC 13139-1851&   15$\pm$4 (F)               &            23$\pm$10  & R2 &  Single \\
$\ldots$ & EC 20140-6935&  -24$\pm$2 (U)               &            17$\pm$10 & R2 &  Binary $^{f}$ \\
$\ldots$ & PB 5418      &  147$\pm$3 (U)               &           152$\pm$10 & R2 &  Single \\
$\ldots$ & PHL 159      &   87$\pm$2 (U)               &            88$\pm$3  & R2 &  Single \\

\hline
\end{tabular}
\flushleft {
Note: $^{a}$ Spectrograph used for the observation: F$=$Feros, T$=$Tull and U$=$UVES  \\
$^{b}$ References: R1=Gontcharov (2006), R2=Silva \& Napiwotzki (2011), R3=Kharchenko et al. (2003), R4=Martin (2006), R5=Kordopartis et al. (2013), R6=\citet{kil1975}  \\
$^{c}$ Martin (2006) obtained the velocity 29$\pm$6 km s$^{-1}$ and quotes also $-15$ km s$^{-1}$ from Duflot et al. (1998).  \\
$^{d}$ R2 gives the velocity as 84 km s$^{-1}$   \\
$^{e}$ R4 also gives velocities of 24.9$\pm$1.5 km s$^{-1}$ from \citet{barbier2000} and 26.5$\pm$2.4 km s$^{-1}$ from \citet{behr2003}.  \\
$^{f}$ See text
}

 \end{table*}

Radial velocities  are provided in Table \ref{stars_rv}. Of the eight stars observed twice or thrice  by us, 
seven appear to have a constant velocity.  The star with a variable radial velocity is HD 204076 which is certainly a spectroscopic binary; the UVES  but not the FEROS spectrum showed double lines.   HD 219188 is possibly a variable. There is a 25 km s$^{-1}$ difference between our two measurements and online catalogs give  a velocity either close to the mean of
our two or a less positive velocity: 84 km s$^{-1}$ according to Silva \& Napiwotzki (2011) or
64 km s$^{-1}$ according to \citet{gont2006} and \citet{khar2007}.

Inspection of Table \ref{stars_rv} suggests that four stars observed once by us may be binaries on account
of a difference with radial velocities reported in the literature. EC 20140-6935 (HD 192273)  was noted as
a possible binary by \citet{magee1998} who found a 45 km s$^{-1}$ velocity difference between their measurement
and that reported by Rolleston et al. (1997).  Silva \& Napiwotzki (2011) give the velocity as $+17$ km s$^{-1}$
from Rolleston et al. Our velocity is in good agreement with that by Magee et al. HD 188618 was suspected
by Martin (2006) to be a binary  from the velocity difference between his measurement of 29$\pm6$ km s$^{-1}$ and a
previous measurement of $-15$ km s$^{-1}$ by \citet{duflot1998}. Our $-46$ km s$^{-1}$ extends the
velocity range. HD 179407's present and previous velocity measurements barely satisfy our 20 km s$^{-1}$
condition. HIP 70205 (LP 857-24) exhibits a 300 km s$^{-1}$ difference but the only previous measurement is from the
RAVE survey \citep{kordo2013} which is possibly ill-suited to velocity measurements of B stars which provide
few lines in the RAVE bandpass. Obviously, this star deserves further attention. Finally, there is HD 114569
which has no previous velocity measurement.

It seems fair to conclude that our sample contains few spectroscopic binaries and certainly fewer than the 
approximately 30\%  provided by a sample of field B stars in the Galactic disk (Abt \& Levy 1984). Our sample is but very slightly biassed
by the exclusion of known double-lined or even single-lined spectroscopic binaries.  HD 1999 and the double-lined eclipsing 
star HD 138503 (Martin 2003) are the only binary stars observed but not in Table \ref{stars_rv}.  With present understanding of 
the formation of single and double runaway stars by the BSS and CES, it is not possible to interpret the low fraction of 
spectroscopic binaries as a pointer to the more important formation mechanism. 

\section{Concluding remarks}

Our sample of  runaway stars is dominated by B stars which have been ejected from sites of recent star formation in
the Galactic disk. Our non-LTE analysis of their C, N, Mg and Si abundances  and published analyses of these
elemental abundances by the same non-LTE procedures in B stars in three open clusters in the Galactic disk  show
no abundance anomalies among the runaways that may be attributed to either of the leading two mechanisms
(BSS and CES) capable obviously of producing runaways.  The spread in abundances over a range of about 1 dex surely reflects the range in
the Galactocentric distance of the birthplaces of the stars before they are ejected into the halo and the presence of an abundance gradient
in the Galaxy. Consideration of either the projected rotational velocities or the radial velocities is presently
unable to provide a determination of the relative probabilities of the BSS and CES in providing the runaway B stars. Yet, an intensive search for binaries among the B runaway population may yet shed light on the relative contributions of the BSS and CES. 
Ability to identify runaway stars formed by the BSS will be enhanced by extending the non-LTE analysis  to other elements, notably
O and S which appear to have large excesses in [X/Fe]  in the (few) LMXB secondaries analysed for this pair of elements.

\acknowledgements

We thank R. Napiwotzki, the referee, for constructive reports.
DLL thanks A.B.S. Reddy for his help in preparing the manuscript for submission.
CMM is grateful to the Department of Education and Learning (DEL) in Northern Ireland and Queen's University Belfast for the award of a research studentship.  This work was supported by the Science and Technology Faculty Council and the UIC. DLL thanks the Robert A. Welch Foundation of Houston, Texas for support through grant F-634. This work was partly funded by the Director General Discretionary fund at ESO. ESO programme IDs 091.D-0061(A) for FEROS and 093.D-0302(A) for UVES.


\end{document}